\documentclass{article}

\usepackage{arxiv}

\usepackage{color}
\usepackage{multirow}
\usepackage{amstext}
\usepackage{subfigure}
\usepackage{amsfonts}

\usepackage[utf8]{inputenc} 
\usepackage[T1]{fontenc}    
\usepackage{hyperref}       
\usepackage{url}            
\usepackage{booktabs}       
\usepackage{amsfonts}       
\usepackage{nicefrac}       
\usepackage{microtype}      
\usepackage{lipsum}
\usepackage{graphicx}
\graphicspath{ {./images/} }

\title{Towards Emotion-Aware User Simulator for Task-Oriented Dialogue}

\author{
 Rui Zhang \\
  School of Software Engineering\\
  South China University of Technology\\
  Guangzhou, Guangdong, China \\
  \texttt{zhang1rui4@outlook.com} \\
   \And
 Kai Yin \\
  School of Software Engineering\\
  South China University of Technology\\
  Guangzhou, Guangdong, China \\
  \texttt{sekaiyin@mail.scut.edu.cn} \\
   \And
 Li Li \\
  School of Software Engineering\\
  South China University of Technology\\
  Guangzhou, Guangdong, China \\
  \texttt{lili961028@gmail.com} \\

}

\begin{document}
\maketitle
\begin{abstract}
The performance of a task-completion dialogue agent usually affects the user experience: when the conversation system yields an unreasonable response, users may feel dissatisfied. Besides, early termination often occurs in disappointing conversations. However, existing off-the-shelf user simulators generally assume an ideal and cooperative user, which is somewhat different from a real user, and inevitably lead to a sub-optimal dialogue policy. In this paper, we propose an emotion-aware user simulation framework for task-oriented dialogue, which is based on the OCC emotion model to update user emotions and drive user actions, to generate simulated behaviors that more similar to real users. We present a linear implementation\footnote{The source code will be released soon.} that is easy to understand and extend, and evaluate it on two domain-specific datasets. The experimental results show that the emotional simulation results of our proposed framework conform to common sense and have good versatility for different domains. Meanwhile, our framework provides us with another perspective to understand the improvement process of the dialogue policy model based on reinforcement learning.
\end{abstract}


\section{Introduction}

The task-oriented dialogue system aims to assist human users in completing domain-specific tasks through human-computer interaction. Generally speaking, such a conversational system consists of the following parts. The Natural Language Understanding (NLU) module maps free text into structured information representations. The Dialogue State Tracking (DST) module records the evolving state of the conversation, and it usually requires collaboration with the Knowledge Base (KB). The Dialogue Policy (DP) model selects appropriate system reactions based on the current state of the dialogue. The Natural Language Generation (NLG) module maps the actions generated by the dialogue agent to human-understandable natural language texts.

For the core component Dialogue Policy model, traditional dialogue systems might be programmed explicitly with manually constructed rules. However, designing reasonable policy rules for complex tasks may result in high costs. Moreover, as the user's behavior and emotional state change, the best policy might be different, which further increases the difficulty of rule design. Therefore, the automatic construction of dialogue policy through reinforcement learning has attracted the attention of academia and industry in recent years.

\subsection{Why Is Emotion-Aware User Simulator Needed?}
Task-oriented dialogue policy optimization can usually be regarded as supervised learning (SL) or reinforcement learning (RL) tasks. In SL-based methods, a policy model is trained to mimic the behavior of experts. This method usually requires a large amount of data labeled by domain experts for training. Therefore, for a task-specific domain, costly and time-consuming data collection and annotation are often required. Moreover, the SL-based method lacks the ability to explore the unknown (dialogue) state space, which limits the ability of the policy model to find the optimal dialogue strategy.

On the contrary, for the RL-based method, the agent can improve the dialogue policy according to the reward signals from the environment without any expert-generated examples. Unfortunately, RL-based methods require a large number of human-computer interaction samples for model optimization, which it too costly and impractical, especially when the policy model is cold-started. To overcome this problem, many researchers apply user simulators to train their RL-based dialogue agents. The purpose of user simulation is to generate natural and reasonable dialogue so that RL agents can explore and learn from the trajectories that may not exist in collected observation data, thereby overcoming the main limitations of SL-based method. Subsequently, these policy model trained with simulated experience can be applied as a reasonable starting point for further improvement through reinforcemet learning in a real dialogue environment.

However, the difference between the user simulator and the real user may cause a suboptimal dialogue policy. In particular, the user simulator plays as an ideal user in most cases: the user simulator will always patiently respond to all the requests from the policy model. Therefore, the user simulator cannot generate any clear negative rewards (penalties) for the agent's apparently unreasonable behavior (for example, repeatedly enquire the known information). In contrast, human users will express dissatisfaction in the conversation, and a common way to achieve this is to terminate the unbearable dialogue as early as possible. This difference is particularly obvious in the learning curve: RL-based dialogue policy model trained with user simulator always report convergence earlier than the one trained in a human-in-loop manner, but this advantages are not shown in manual evaluation. In other words, how similar the user simulator is to a human user largely determines the performance of the trained policy model in a real environment. Therefore, instead of assuming an ideal person, we hope to introduce emotional features to build a more realistic user simulator framework. Besides, another advantage of constructing an emotion-aware user simulator is that by simulating the user's emotional state, we are allowed to use emotional information as a clue for policy optimization, which is rarely studied in the current RL-based policy learning methods.

\subsection{Background}
\subsubsection{Emotional Task-Completion Dialogue}

Previous research has shown that human users usually impose their own emotional and interpersonal patterns on computer systems, and expect the dialogue agents to respond in an empathetic manner\cite{prendinger2003using,reeves1996media}. As early as 2000, Polzin's pioneering work \cite{polzin2000emotion}, which focused on switching response strategies based on the user's emotional state, first explored this problem. Andre \cite{andre2004endowing} combined the social theory of politeness with the cognitive theory of emotion, and proposed a hierarchical selection process to manifold the generation style of the dialogue system. Skowron \cite{skowron2010affect,skowron2011good}proposed an Affect-Listener framework that responds to the user's utterance at both content- and emotion- levels.

In these early works, the task-completion part of the dialogue policy is usually implemented in a rule-based method or a Partially Observable Markov Decision Process (POMDP) method. In the new wave of deep learning (DL), many task-oriented dialogue policy models based on neural networks have been investigated to help users achieve their goals \cite{gao2019neural}, while there is still a lack of relevant research involving user emotions. One of the main reasons for this phenomenon is that DL-based approaches usually require a large amount of data for model learning, but the existing publicly labeled task-oriented dialogue data usually not contain emotional information. Meanwhile, the current off-the-shelf user simulators also ignore the simulation of user emotions, which makes it difficult to effectively research task-oriented dialogue policy learning methods involving user emotions.

%

\subsubsection{User Simulator}

Training a dialogue policy model through direct reinforcement learning is always costly, as it requires a lot of interactions with real users. The user simulator, as a cheaper alternative, has been widely used in reinforcement learning based dialogue policy optimization \cite{keizer2012user,shi2019build}. 

Most user simulators are rule-based because they are generally easy to understand and extend. Schatzmann et al. proposed an agenda-based method for simulating user behaviours \cite{schatzmann2007agenda}. Li et al. \cite{li2016user} proposed an easy-to-use rule-based user simulator for ticket booking dialogue.
However, rule-based user simulators usually require a lot of manual maintenance costs. In recent years, data-driven user simulators \cite{khouzaimi2017incremental} are also emerging. El et al. \cite{el2016sequence} investigated user simulator based on sequence-to-sequence framework, while Hou et al. \cite{hou2019corpus} presented a state2seq user simulator for task-completion conversation.

Although the discrepancies between the simulator and the real users might result in a sub-optimal agent \cite{peng2018deep}, user simulator is still a useful tool that allows us to analyze and reproduce the experimental results without much cost.


\section{Task-oriented Dialogue Systems}
We consider a dialogue system that can help users achieve tasks in a specific domain through natural language interaction. During the conversation, the agent collects information relevant to the user's goal and assists the user in accomplishing his/her goal. In this paper, we focus on building a user simulator to imitate human users' actions in the above interaction process.

\subsection{Data} We construct our emotion-aware user simulator framework based on the experimental environment provided by \cite{li2017end,wang2019dialogue}. The dataset is composed of dialogue in several domains, and our evaluation is perform on two of them: {\it movie-ticket booking} and {\it taxi ordering}. Each domain has its domain-specific intents and slots, and the dataset contains raw conversational data collected via Amazon Mechanical Turk, and has been manually annotated according to a schema defined by domain experts, as illustrated in Table \ref{table:schema}.
For the movie-ticket booking task, this dataset contains 2890 labeled conversations, while for the taxi-ordering task, the dataset contains 3094 dialogues. 

\begin{table*}[htbp!]
	\setlength{\tabcolsep}{3pt}
	\centering
	\caption{Annotation schema for the dataset. Among them, all intents and some slots are shared in the two domains, and each domain contains its domain-specific slots. And we also add an additional intent "terminating" to represent the early suspension behavior in the conversation.}	\label{table:schema}
	\begin{tabular}{c c c}
		\hline
		Domain & Intent & Slots \\
		\hline
		\multirow{3}{*}{(Shared)} & request, inform, deny, greeting & date, city, zip, state, closing,\\
		& comfirm\_question, confirm\_answer, & distance\_constraints, number\_of\_people, \\
		& multiple\_choice, thanks, closing &  greeting, task\_complete, other \\
		\cline{2-3}
		\multirow{2}{*}{Movie} & \multirow{2}{*}{-} & moviename, price, starttime, theater, \\
		& & ticket, theaterchain, video\_format \\
		\cline{2-3}
		\multirow{4}{*}{Taxi} & \multirow{4}{*}{-} & taxi, dropoff\_location, cost, \\
		& & pickup\_location, taxi\_company,\\
		& & pickup\_city, pickup\_time, \\
		& & dropoff\_city, car\_type, name \\
		\cline{2-3}
		(Extended) & terminating & - \\
		\hline
	\end{tabular}
\end{table*}

\section{The Emotion-Aware User Simulator}
Unlike existing off-the-shelf user simulators, our framework mimic human users on two aspects: task-completion and emotional state. In terms of task-completion, we follow the implementation of \cite{schatzmann2009hidden} and \cite{li2016user}, by using an agenda-based method to simulate the user collaboration to achieve the goals. In terms of emotional state, we follow the basic idea of the OCC psychological model to implement an event-driven emotional simulation method.

The task-completion module receives dialogue state and selects a reasonable response action. The emotion simulation module updates the simulated emotion according to the behavior of the agent and may affect the response action in some cases. In the following section we provide a brief introduction of the task-completion module, and present the detailed implementation of the emotion simulation module.
\subsection{Task-Completion Module}
The task-completion module largely follows the design of \cite{li2016user}, which contains three core sub-modules: user goal generation, natural language understanding and action selection. 

At the outset of a conversation, the simulated user will sample a user goal from the dataset as the current target. The user goal usually includes two parts: (1) inform\_slots: contain a number of slot-value pairs which serve as user constraints, and (2) request\_slots: contain a group of slots that user has no clue about their information but wants to obtain the slots from the conversation.

For the natural language understanding sub-module, the user simulator uses a long-short term memory (LSTM) network to do intent prediction and slot filling simultaneously, as introduced in \cite{hakkani2016multi}.

In each turn of a conversation, a set of manual rules are applied to select the next appropriate action based on the current dialogue state and the user goal to promote the completion of the user target.

\subsection{Emotion Simulation Module}

The emotion simulation module in our user simulator contains three core procedures: (1) {\bf Trigger Detection}: which applies a specific algorithm to detect events in the conversations that may cause the user's emotional changes (namely {\it triggers}); (2) {\bf Emotional State Updating}: which alters user's current emotional state according to the user's personality, historical emotional state and the current emotional change; and (3) {\bf Action Driving}: which drives the user behavior based on the current dialogue state and user emotional state.

In the following sections, we will discuss the details of the emotion module in our proposed emotion-aware user simulator, and present a linear implementation that is easy to understand and extend.

\subsubsection{Trigger Detection}

The Trigger Detection procedure simulates the real user's perception of events in the external environment, which may involve acoustic, optical and textual information. In this paper, we mainly consider the events caused by the dialogue system, which are usually detected using text features. According to the OCC theory \cite{ortony1988cognitive}, events in the environment greatly affect the type and intensity of emotions. Thus, the goal of the Trigger Detection procedure is to identify these events from the given conversation context and generate the corresponding emotional variation.

Generally speaking, the emotional trigger events in the task-completion conversations depend on the specific task domain. Except for certain mutually exclusive cases, we assume that all these emotional triggers can occur simultaneously, including but not limited to:
\begin{itemize}
	\item {\bf Overlong dialogue.} The purpose of task-oriented dialogue is to help users complete specific tasks through conversations with limited length. Dialogues that are overlong and tedious can easily cause user dissatisfaction, where the acceptible length varies according to the complexity of the task in particular domain.
	\item {\bf Irrelevant/relevant response.} If the system's response is irrelevant to the current user's goal, the user experience will decline; on the contrary, if the system is able to respond in a timely and reasonable manner, user satisfaction will increase.
	\item {\bf Repeated query.} Users' discontent may increase if the conversational agent keeps asking the same information that has already been informed by the user.
	\item {\bf Initiative.} A well-designed dialogue system can actively provide information related to the current task instead of waiting for the user to ask all the information item by item. Such initiative helps improve the user experience.
	\item {\bf Suggestion.} In some task-completion dialogue, when the available resources cannot meet the user's needs or constraints, the agent will provide users with alternative options. Such suggestions may cause changes in the user's emotion, depending on its rationality.
\end{itemize}

In addition, according to the OCC theory, the same event has different emotional impact on individuals with different personalities. In task-completion dialogue, we assume that all users want to achieve their user goal, but expect that the differences in personality between users can be reflected in emotional changes. Therefore, in our trigger detection procedure, the emotional variation depends on both the event and the personality of the user, which is formalized as:

\begin{equation}
v_t = D(\mathbf{\hat{T}}_t, \mathbf{p}) 
\end{equation}
where $\mathbf{\hat{T}}_t$ indicates the detected triggers at time $t$, $\mathbf{p}$ represents the user's personality, and $v_t$ is the corresponding emotional variation.

\subsubsection{Emotional State Updating}
The emotional state updating module is proposed to simulate the user's sentimental change process. Unlike the discrete dialogue states, the emotional states of a user in a conversation are usually time-sequential: Emotions usually change gradually, and the current emotional state is always affected by the previous one. Therefore, the emotional state updating process needs to consider user's historical emotional information, in which the updating procedure can be regarded as composed of external and internal changes. Among them, external changes are caused by environmental or event-related stimuli, while internal changes refer to the process in which emotions gradually weaken and disappear over time.

In this paper, we adopt an emotional update process similar to the previous work \cite{egges2004generic}. Given the previous emotional state $e_{t-1}$, the historical emotional states $e_{1:t-1}$, the user's personality $\mathbf{p}$ and the emotional variation $v_t$ yield by the trigger detection process, the current emotional state $e_{t}$ is calculated as:
\begin{equation}
e_{t} = e_{t-1} + \Theta(e_{1:t-1}, \mathbf{p}, v_t) + \Phi(e_{t-1}, \mathbf{p})	\label{formula:etp1}
\end{equation}
where $\Theta(e_{1:t-1}, \mathbf{p}, v_t)$ is an update function which calculates the actual change of the user's emotion according to the historical emotional states $e_{1:t-1}$, the personality $p$ and the current emotional variation $v_t$. And $\Phi(e_{t-1}, \mathbf{p})$ is a decay function based on both $e_{t-1}$ and $\mathbf{p}$.

\subsubsection{Action Driving}
Although the influence of emotion on human behavior has been confirmed by many previous studies \cite{dobson2010historical} in the field of psychology, its internal influence mechanism is still a subject to be further explored. In task-oriented conversation, we are more concerned about how the user and the dialogue system collaborate to complete a specific task, and how the suspension of the conversation is caused when the user is dissatisfied. 

Thus, a simple conditional probability is used to describe this process in our framework: we assume that users will terminate the conversation with a certain termination probability $p_{\text{term}}$ in advance when the intensity of their negative emotions is higher than a certain threshold $\eta_b$. Among them, $\eta_b$ and $p_{\text{term}}$ can either be fixed values manually set, or be varying values on interval constructed by a continuous function.

\section{A Linear Implementation}
In this section, we provide a linear implementation of the emotional module, which is simple yet effective and can be easily extended, to illustrate how emotion is simulated in our proposed framework. 


\subsection{Personality, Emotion, and Events}
We assume that the personality of user can be divided into five dimensions: \{{\it Openness, Conscientiousness, Extraversion, Agreeableness, Neuroticism}\}, according to the Big Five personality theory \cite{digman1990personality,mccrae1992introduction}. For each user we use a one-dimensional vector:
\begin{equation}
\mathbf{p} = \{p^{\text{Open}}, p^{\text{Cons}}, p^{\text{Extra}}, p^{\text{Agree}}, p^{\text{Neuro}}\}
\end{equation}
where each $p^* \in \left[0, 1\right]$ represents the weight of each personality dimension. 

Besides, six kinds of emotions \{{\it Angry, Disgust, Fear, Happiness, Sadness}, and {\it Surprise}\} are defined in our implementation following the emotional theory proposed by Ekman \cite{ekman2013emotion}. The user's emotional state at time $t$ can be represented as a one-dimensional vector:
\begin{equation}
e_t = \{e_t^{\text{Ang}}, e_t^{\text{Dis}}, e_t^{\text{Fear}}, e_t^{\text{Hap}}, e_t^{\text{Sad}}, e_t^{\text{Sur}}\}
\end{equation}
where each $e_t^* \in [0, 1]$ represents the emotional intensity of the corresponding category at time $t$.

We implement some rule-based detection methods for five aforementioned common triggers \{{\it Overlong Dialogue, Irrelevant/Relevant Response, Repeated Query, Initiative}\}, including:
\begin{itemize}
	\item We define a conversation to be {\it overlong} if it exceeds the length $\tau$, where $\tau$ is a threshold manually determined according to the domain and task complexity.
	\item If the dialogue system mentions some information (slots) that is entirely unrelated to the current user's goal, we treat it as an irrelevant response; otherwise, it is a relevant response.
	\item A query is considered as repeated if the information it provides has been informed in the previous conversation.
	\item We define initiative as the behavior of a dialogue system that actively informs users of the information they want to request.
\end{itemize} 

At each time step $t$, we detect the above events caused by the current conversation, and represent them as a one-dimensional vector:
\begin{equation}
\mathbf{\hat{T}}_t = \{T_t^{\text{Od}}, T_t^{\text{Ir}}, T_t^{\text{Rr}}, T_t^{\text{Rq}}, T_t^{\text{In}}\}
\end{equation}
where $T_t^* \in \{0, 1\}$ represents whether the corresponding event happens at time step $t$.

\subsection{Calculation of Emotional Variation}
The trigger detection procedure first applies the aforementioned rule-based method to identify the triggering events $\mathbf{\hat{T}}_t$ in the conversation at time step $t$. By maintaining an trigger-emotion matrix $M_{te}$ which indicates the intensity of emotional variant that each trigger will produce, the emotional variant can be calculated as:
\begin{equation}
v_t = \mathbf{\hat{T}}_t \cdot M_{te} 	\label{formula:vt_simple}
\end{equation}
where $M_{te} \in \mathbb{R}^{|T| \times |E|}$ is a weighted matrix, with $|T|$ be the total number of triggers and $|E|$ be the number of emotional categories ($|T|=5, |E|=6$ in our case).

Nevertheless, our model assumes that users with specific personalities have different sensitivity to various triggering events. To this end, we apply a personality-trigger matrix $M_{pt}$ to reflect how users with different personalities pay attention to triggers. Therefore, unlike Formula \ref{formula:vt_simple}, for a user with personality $\mathbf{p}$, the amount of emotional variant produced by event $\mathbf{\hat{T}}_t$ should be calculated as:
\begin{equation}
v_t = ((\mathbf{p} \cdot M_{pt})*\mathbf{\hat{T}}_t) \cdot M_{te}
\end{equation}
where $M_{pt} \in \mathbb{R}^{|P| \times |T|}$ is a weighted matrix, with $|P|$ be the number of personality factors ($|P|=5$ in our case). 

\subsection{Update the Simulated Emotional State}
According to Formula \ref{formula:etp1}, the change of user emotional state can be regarded as the joint effect of the update procedure $\Theta(e_{1:t-1}, \mathbf{p}, v_t)$ and the decay procedure $\Phi(e_{t-1}, \mathbf{p})$. We present a linear implementation of $\Theta(e_{1:t-1}, \mathbf{p}, v_t)$ similar to \cite{egges2004generic} in this sample, calculated as:
\begin{equation}
\Theta(e_{1:t-1}, \mathbf{p}, v_t) = \Theta(\mathbf{p}, v_t) =(\mathbf{p} \cdot M_{pe}) * v_t
\end{equation}
where $M_{pe} \in \mathbb{R}^{|P| \times |E|}$ calculates the importance of each emotion depending on the personality dimensions. As can be noted, this implementation makes no use of the historical emotional state $e_{1:t-1}$. The decay function $\Phi(e_{t-1}, \mathbf{p})$ is designed as a model with constant decay weight regardless of the personality in our presented linear implementation, calculated as:
\begin{equation}
\Phi(e_{t-1}, \mathbf{p}) = \{-C^1,\cdots, -C^{|E|} \} * e_{t-1}
\end{equation}
where $C^i \in [0, 1]$ defines the amount of decay for each emotion.

\subsection{Emotion-Motivated Behavior}
We adopt a simple rule to simulate the early termination of the conversation: When the overall intensity of the user's negative emotions exceeds his/her positive emotions, the user will terminate the current conversation with a probability $p_{\text{term}}$. We first normalize the emotional intensity $e_t$ so that $\sum_i e_t^i = 1$. When $\text{sum}(e_t^{\text{Ang}}, e_t^{\text{Dis}}, e_t^{\text{Fear}}, e_t^{\text{Sad}}) > \eta_b$, the user simulator yields a "terminating" action to early suspend the current dialogue, where the termination threshold $\eta_b = 0.5$; else, the user simulator generate the current action according to the result of the task-completion module.

\section{Model Performance}

\subsection{Basic Features}	\label{section:sim_exp}
We conduct experiments on two domain-specific tasks to evaluate whether the proposed framework meets the design requirements. Our experiment contains two different personality groups (denoted as $u_\text{A}$ and $u_\text{B}$) to evaluate the simulator's ability to mimic individuals with different personalities. In addition, we provide a group that allow termination (where $p_{\text{term}}=0$) to study the impact of early termination on the dialogue process. Table \ref{table:result}, Figure \ref{fig:draw1} and \ref{fig:draw2} in the Appendix demonstrate the results of the simulation experiment. Since in the field of Movie-ticket booking and Taxi-ordering, user emotions rarely involve the two categories \{{\it Fear} and {\it Sadness}\}, we only report four emotional categories \{{\it Angry, Disgust, Happy} and {\it Surprise}\}.

It can be seen that the linear implementation we proposed satisfies the following basic features, which are detailedly discussed and analyzed in the appendix:
\begin{itemize}
	\item Different individuals will lead to different simulated emotional state given the same dialogue trajectory. 
	\item Different settings of termination probability $p_{\text{term}}$ will cause different dialogue process. 
	\item Our framework can be easily extended for task-oriented dialogues in different domains. 
\end{itemize}

\subsection{Implementation Details}	\label{section:man_exp}
In our linear implementation, several weight matrices ($M_{te}, M_{pt}$ and $M_{pe}$) are used to represent the relationship between events, personalities and emotions. However, the weights of these matrices need to be set based on human experience, while the research work involved in this aspect is still scarce. In this paper, we adopt a two-step initialization method. In the first step, we first assign posible initial values to these weight matrices. In the second step, we recruit 15 volunteers to interact with the dialogue system and record the conversation experience for model weight adjustment. Each volunteer was asked to take a test to evaluate their Big-Five Personality factors at the outset. In each dialogue turn, the volunteers are asked to record their current emotional state and intensity (ranged from level 1 to 5). For each volunteer, we collect up to 50 dialogues experiences. Finally, we manually adjust the previous matrix weights according to the collected experience to better reflect the emotional changes of real users.

\section{Discussion}	\label{section:discuss}
According to the results on Figure \ref{fig:draw1} and \ref{fig:draw2}, our proposed framework can effectively simulate user emotions. As the success rate of dialogue increases, the simulated intensity of the happiness increase, while anger and disgust decrease accordingly. At the same time, the proportion of events that would cause lower user experience decreases significantly over time, while the proportion of positive events (such as RR and IN) gradually increases. This is consistent with our intuitive understanding of a task-oriented dialogue policy optimization process.

In short, the {\bf main advantage} of our proposed framework include:
\begin{itemize}
	\item Compared with the existing simulators that assumes an ideal user, our framework can produce simulated behaviours that are more similar to real user behaviours, thereby reducing the suboptimality of the dialog policy model. On the one hand, for individuals with different personalities, our framework sill simulate more complex emotional states and changes. This helps to improve the generalization ability of the policy model, and allows researchers to customize corresponding dialogue policies for user groups with different personalities. On the other hand, our emotion-aware user simulation can imitate the user's early suspension behavior in a conversation, instead of simulating the user as an ideal individual who always collaborates. This allows the policy model to be trained and evaluated in a simulation framework closer to the real environment.
	\item The emotional simulation module in our framework has good domain adaptability, according to the results of the simulated experiment and manual evaluation. In our framework, the emotional changes are driven by events which are independent of specific domain knowledge and can be effectively identified through textual and dialogue state cues. Moreover, these events are ubiquitous in task-oriented conversation of different domain. Therefore, our emotional simulation module can be easily transferred to different fields with very few modifications.
	\item In addition, our emotion-aware user simulator also provides us with a novel perspective to evaluate dialogue policy in a reproducible way. In particular, dialogue policy optimization based on deep Q-learning is usually a black box process, which prevents researchers from knowing whether the model is being optimized, especially when the success rate curve is flat in the intermediary stage of the model training procedure. Through our simulator framework, we can further observe and analyze the optimization process of the dialogue policy from the perspective of events and emotions: whether the repetitive/irrelevant responses in the conversation are reduced, whether the user experience is improved, and so on.
	
\end{itemize}

However, there are still some details that {\bf need improvement} in future work:

\begin{itemize}
	\item Most of the weight matrices are still manually set based on experience in our linear implementation. Although we have optimized and adjusted these values based on the experience of real users through human-in-loop experiment, due to the limited size of samples, these hyper-parameters are still not enough to completely model the real emotional change process.
	\item Some more complex events in task-completion dialogue have not been taken into consideration in our linear implementation. For example, some dialogue systems allow the agent to proactively provide suggestions when the constraints of the user goal cannot be met. Some cross-domain task scenarios contain more complex constraints and events (such as a intelligent travel assistant that can help users book hotels and air tickets). These tasks undoubtedly increase the difficulty of emotion analysis and simulation, which has not yet been covered by our current implementation.
\end{itemize}

\appendix   
\setcounter{table}{0}   
\setcounter{figure}{0}
\renewcommand{\appendixname}{Appendix~\Alph{section}}
\renewcommand{\thetable}{A\arabic{table}}
\renewcommand{\thefigure}{A\arabic{figure}}

\section{Appendix}
\subsection{Evaluation of Basic Features}

\begin{table*}[htbp!]
	\setlength{\tabcolsep}{6pt}
	\center
	\caption{Results of different settings at training epoch=\{100, 200, 300\}. Each number is average over 5 runs. (OD: Overlong Dialogue, IR: Irrelevant Response, RQ: Repeated Query, RR: Relevant Response, IN: Initiative)}	\label{table:result}
	\begin{tabular}{c c | c c c | c c c}
		\hline
		\multirow{2}{*}{Epoch} & \multirow{2}{*}{Criterion} & Movie & Movie & Movie & Taxi & Taxi & Taxi \\
		& & ($u_\text{A}$) & ($u_\text{A}$+terminate) & ($u_\text{B}$) & ($u_\text{A}$) & ($u_\text{A}$+terminate) & ($u_\text{B}$) \\
		\hline
		& Success & 0.452 & 0.425 & 0.452 & 0.031 & 0.062 & 0.031 \\
		\cline{2-8}
		& Angry & 0.210 & 0.205 & 0.194 & 0.433 & 0.402 & 0.420 \\
		& Disgust & 0.183 & 0.191 & 0.164 & 0.310 & 0.328 & 0.292 \\
		\multirow{2}{*}{Epoch} & Happy & 0.526 & 0.509 & 0.538 & 0.124 & 0.145 & 0.122 \\
		\multirow{2}{*}{100} & Surprise & 0.081 & 0.094 & 0.104 & 0.132 & 0.128 & 0.167 \\
		\cline{2-8}
		& OD & 21.9\% & 22.0\% & 21.9\% & 40.8\% & 36.6\% & 40.8\% \\
		& IR & 6.9\% & 15.3\% & 6.9\% & 25.4\% & 24.8\% & 25.4\% \\
		& RQ & 21.0\% & 19.3\% & 21.0\% & 13.5\% & 18.6\% & 13.5\% \\
		& RR & 34.8\% & 31.8\% & 34.8\% & 14.6\% & 14.8\% & 14.6\% \\
		& IN & 15.4\% & 11.6\% & 15.4\% & 5.7\% & 5.3\% & 5.7\% \\
		\hline
		& Success & 0.611 & 0.574 & 0.611 & 0.283 & 0.214 & 0.283 \\
		\cline{2-8}
		& Angry  & 0.133 & 0.142 & 0.139 & 0.365 & 0.367 & 0.352 \\
		& Disgust & 0.123 & 0.129 & 0.115 & 0.318 & 0.311 & 0.302 \\
		\multirow{2}{*}{Epoch} & Happy & 0.684 & 0.665 & 0.665 & 0.194 & 0.207 & 0.192 \\
		\multirow{2}{*}{200} & Surprise & 0.060 & 0.064 & 0.081 & 0.123 & 0.115 & 0.154 \\
		\cline{2-8}
		& OD & 10.9\% & 12.6\% & 10.9\% & 30.1\% & 30.1\% & 30.1\% \\
		& IR & 3.5\% & 6.8\% & 3.5\% & 20.8\% & 18.5\% & 20.8\% \\
		& RQ & 18.2\% & 14.6\% & 18.2\% & 12.9\% & 12.7\% & 12.9\% \\
		& RR & 40.3\% & 40.8\% & 40.3\% & 23.2\% & 25.2\% & 23.2\% \\
		& IN & 28.8\% & 25.3\% & 28.8\% & 12.9\% & 13.5\% & 12.9\% \\
		\hline
		& Success & 0.632 & 0.590 & 0.632 & 0.332 & 0.324 & 0.332 \\
		\cline{2-8}
		& Angry & 0.111 & 0.127 & 0.128 & 0.371 & 0.369 & 0.359 \\
		& Disgust & 0.113 & 0.134 & 0.111 & 0.280 & 0.299 & 0.263 \\
		\multirow{2}{*}{Epoch} & Happy & 0.721 & 0.679 & 0.684 & 0.228 & 0.210 & 0.226 \\
		\multirow{2}{*}{300} & Surprise & 0.055 & 0.059 & 0.077 & 0.121 & 0.121 & 0.152 \\
		\cline{2-8}
		& OD & 8.3\% & 11.7\% & 8.3\% & 31.0\% & 30.6\% & 31.0\% \\
		& IR & 4.2\% & 7.9\% & 4.2\% & 16.2\% & 18.1\% & 16.2\% \\
		& RQ & 15.3\% & 14.6\% & 15.3\% & 12.8\% & 12.9\% & 12.8\% \\
		& RR  & 43.1\% & 40.5\% & 43.1\% & 25.6\% & 24.7\% & 25.6\% \\
		& IN  & 29.1\% & 25.3\% & 29.1\% & 14.4\% & 13.7\% & 14.4\% \\
		\hline
	\end{tabular}
	
\end{table*}

\begin{figure*}
	\centering
	\subfigure[Result for $u_A$.]{
		\includegraphics[width=0.35\linewidth]{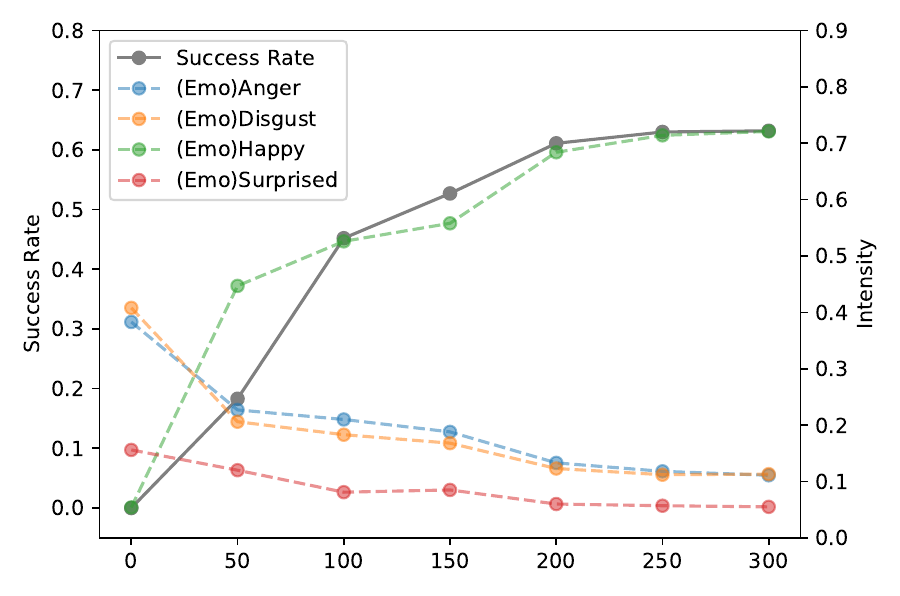}
		\includegraphics[width=0.35\linewidth]{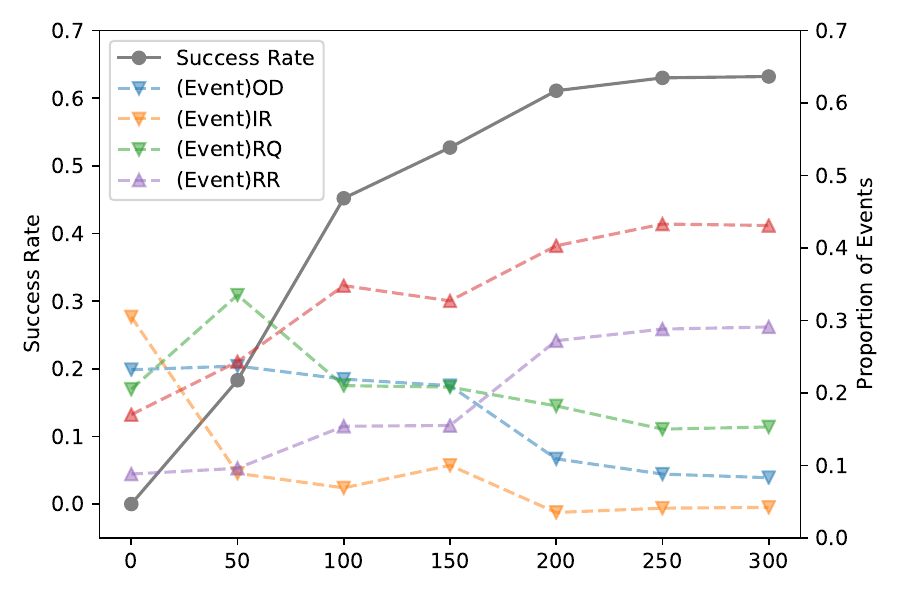}
	}
	\subfigure[Result for $u_A$+terminate.]{
		\includegraphics[width=0.35\textwidth]{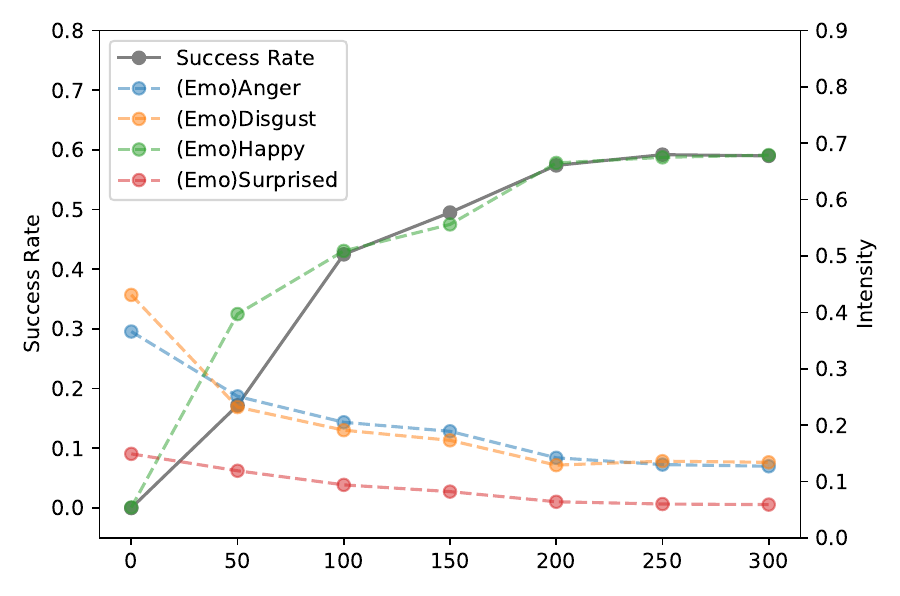}
		\includegraphics[width=0.35\textwidth]{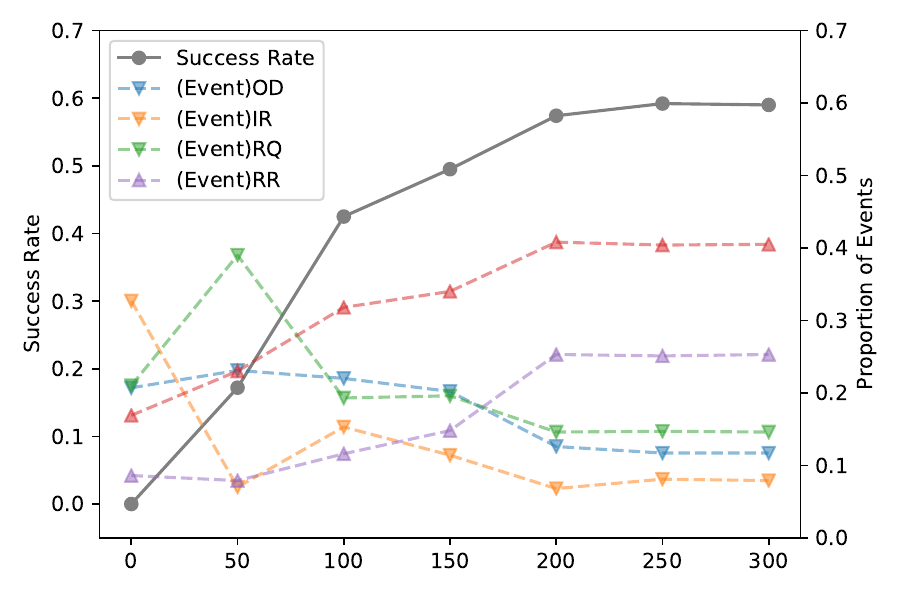}
	}
	\subfigure[Result for $u_B$.]{
		\includegraphics[width=0.35\textwidth]{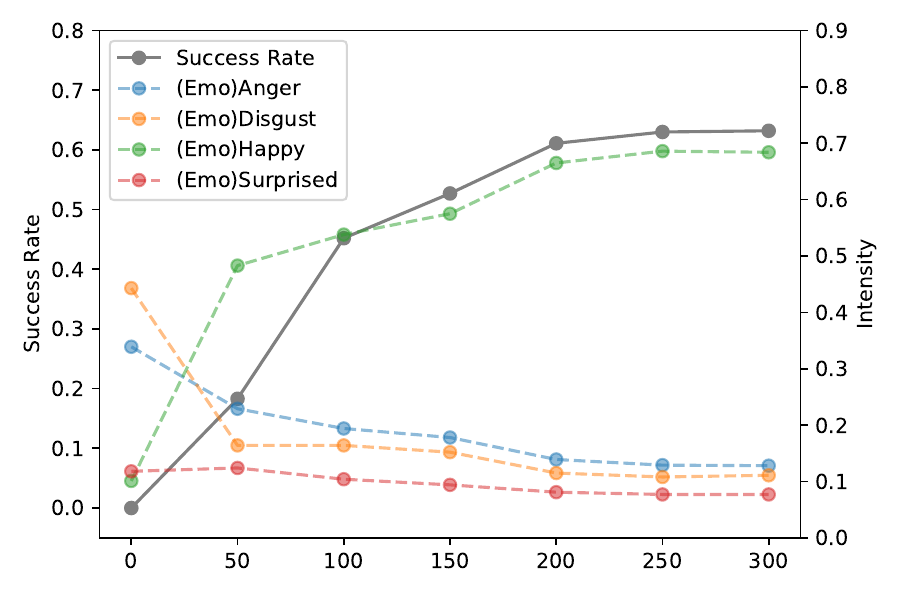}
		\includegraphics[width=0.35\textwidth]{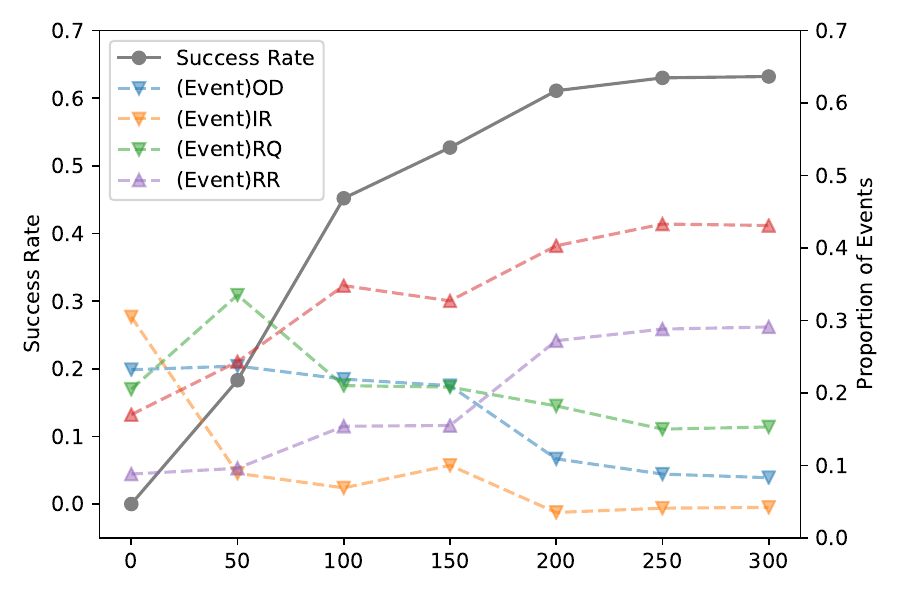}
	}	
	\caption{The simulated emotional intensity (left) and detected trigger events (right) under different personality settings on the movie-ticket booking dataset. The upper triangle mark in the figure indicates that the corresponding attribute contributes to positive user emotional state, while the lower triangle mark indicates that it will lead to negative user emotions.}
	\label{fig:draw1}       
\end{figure*}

\begin{figure*}
	\centering
	\subfigure[Result for $u_A$.]{
		\includegraphics[width=0.35\textwidth]{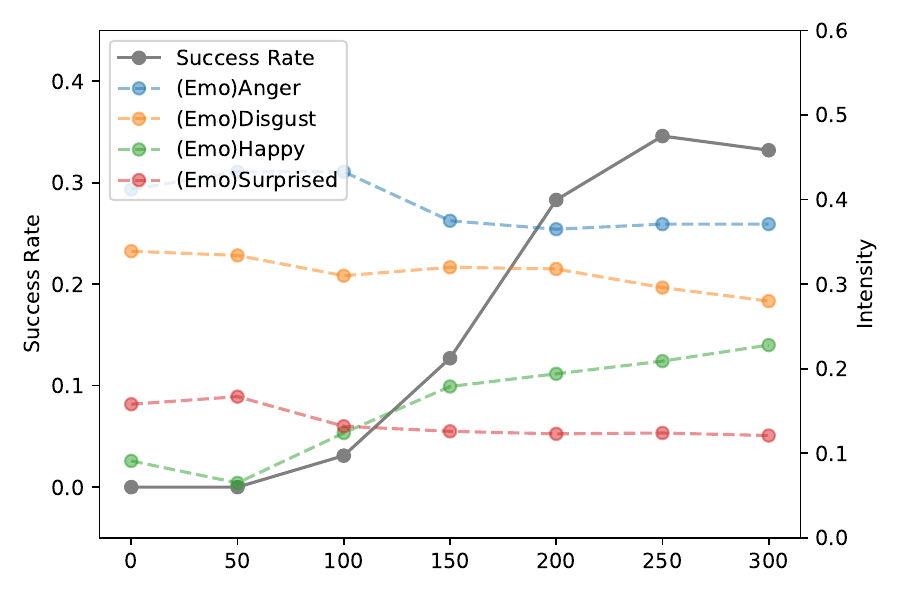}
		\includegraphics[width=0.35\textwidth]{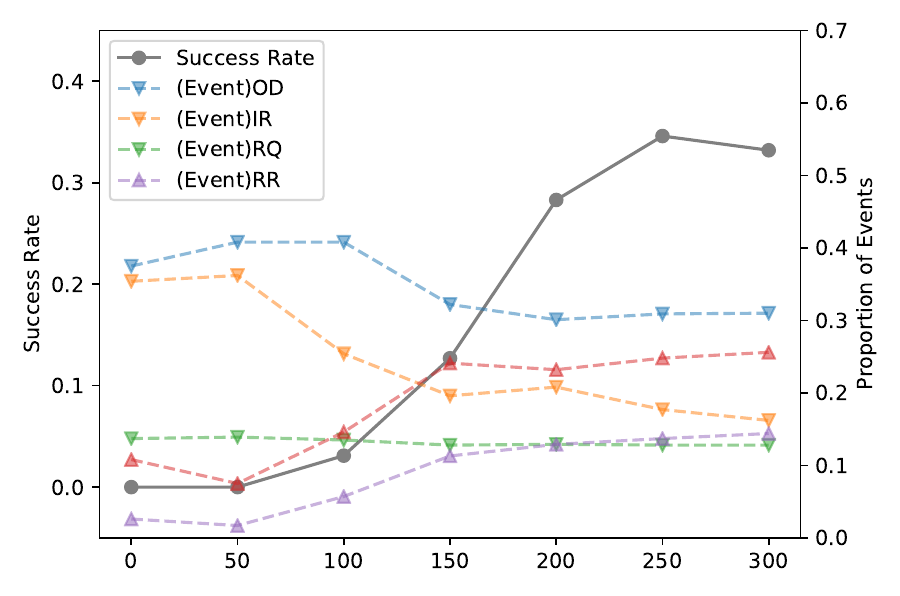}
	}
	\subfigure[Result for $u_A$+terminate.]{
		\includegraphics[width=0.35\textwidth]{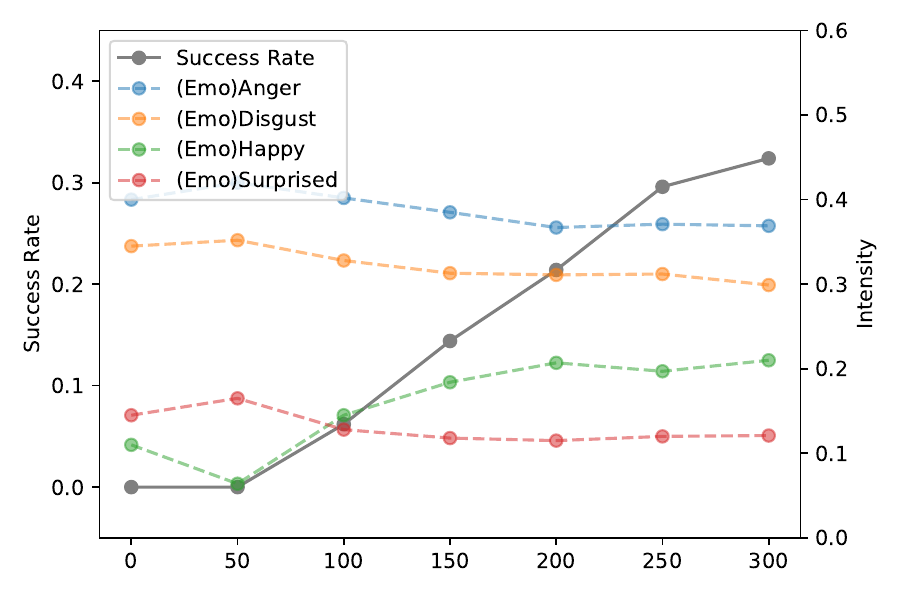}
		\includegraphics[width=0.35\textwidth]{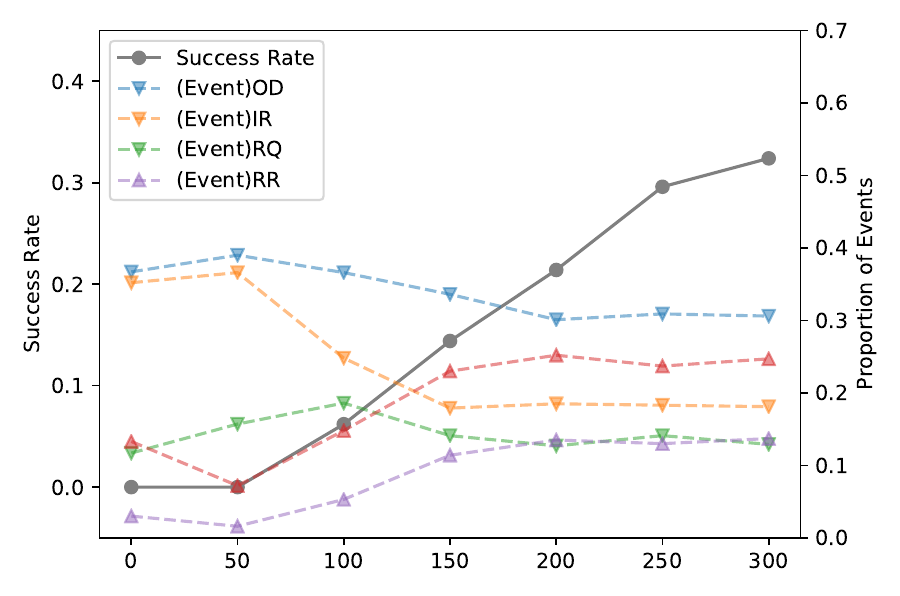}
	}
	\subfigure[Result for $u_B$.]{
		\includegraphics[width=0.35\textwidth]{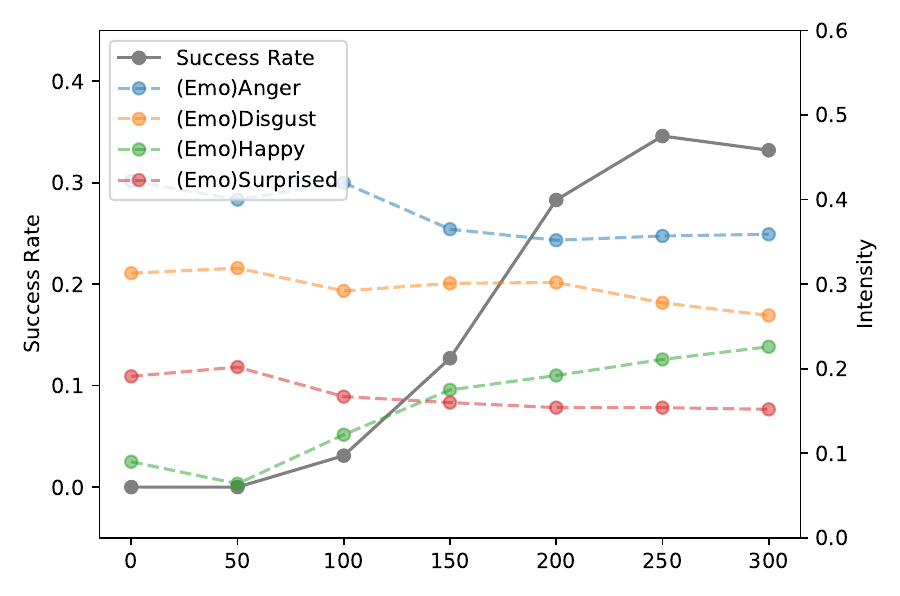}
		\includegraphics[width=0.35\textwidth]{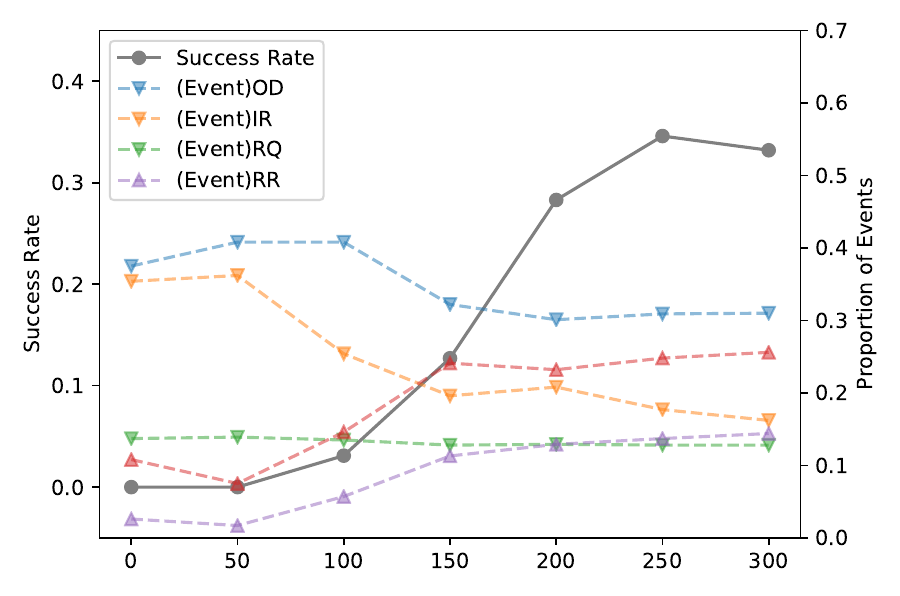}
	}	
	\caption{The simulated emotional intensity (left) and detected trigger events (right) under different personality settings on the taxi-ordering dataset.}
	\label{fig:draw2}       
\end{figure*}

In this section, we briefly introduce how we conduct experiments and evaluate whether our linear implementation meets our design requirements.

We designed two simulated users A and B, where the personality test results of two volunteers were used to initialize the personalities $\mathbf{p}$ of the simulated users. For tasks in the same domain, we designed two experiments: (1) By comparing the simulated emotion of the two users given the same dialogue trajectory (without early termination), we tried to verify that the simulated users with different personalities have diverse reactional emotions. (2) By turning on/off the early termination setting, we tried to explore the impact of the breakdown in conversation on the dialogue trajectory and emotional state of the same simulated user. In addition, we conducted the above evaluations on both the movie-ticket booking dataset and the taxi-ordering dataset, to verify that our framework is universal for different domains.

End-to-end dialogue policy optimization approach is applied in our experiments. The policy model is represented as a standard deep Q-learning network \cite{mnih2015human}, which tasks the dialogue state $s_t$ at time $t$ as input and outputs the $Q(s_t, a)$ values of all actions for action selection. We use common deep Q-learning tricks, such as target network usage and experience replay strategy, to improve DQN performance. The hidden size of the policy network is set to 80.

Table \ref{table:result} shows the results of the above evaluation. All values in the table are based on the average of 5 repeated experiments (using different fixed random seeds).

\subsubsection{Simulation of Different Personalities}
By comparing the column pairs Movie($u_\text{A}$)-Movie($u_\text{B}$) and Taxi($u_\text{A}$)-Taxi($u_\text{B}$) in Table \ref{table:result}, we can find that in these pairs, users A and B have the same success rate, average turns, and number of triggers, but their corresponding emotional intensities are different. This shows that our framework can simulate a variety of emotional experiences according to different personalities. In the above experiment, the termination probability $p_{\text{term}}$ is set to 0, that is, no matter how the conversation is conducted, the user will never terminate the dialogue in advance until the conversation exceeds the maximum length. This ensures that the dialogue process will not be different due to breakdown.
\subsubsection{Early Termination in Dialogue}
By comparing the column pairs Movie($u_\text{A}$)-Movie($u_\text{A}$+terminate) and Taxi($u_\text{A}$)-Taxi($u_\text{A}$+terminate) in Table \ref{table:result}, we can find that when the termination mode is turned on/off, the success rate, average turns and number of triggers for the same user are also different. This is because the user's interruption behavior will cause the conversation to be terminated early, which changes the original trajectory of the conversation process. We show the simulated emotional experience and the distribution of detected trigger events under different $p_{\text{term}}$ settings in Figures \ref{fig:draw1} and \ref{fig:draw2}.
\subsubsection{Model Performance in Different Domain}
Since the trigger events in our proposed simulator are universal for different fields, the cost for adaptation is very low. According to the results of Table \ref{table:result}, by comparing the performance on two domain-specific datasets, it can be seen that the emotional simulation performance of our method is consistent and stable. The results of Figure \ref{fig:draw1} and \ref{fig:draw2} also show that our framework achieves good performance for different task scenarios, as the simulation performance is consistent with our intuitive cognition.

%
%

\bibliographystyle{unsrt}
\bibliography{references}


\end{document}